\documentstyle[preprint,aps]{revtex}
\draft      
\begin{document}
\title{Single Atom and Two Atom Ramsey Interferometry with Quantized Fields}
\author{G. S. Agarwal,$^{1}$ P. K. Pathak $^{1}$ and M. O. Scully $^{2,3}$}
\address{1. Physical Research Laboratory, Navrangpura, Ahmedabad 380 009,
India\\2. Department of Physics, Texas A \& M University, College Station, Texas
77843 \\
3.Max-Planck Institut f$\ddot{u}$r Quantenoptik, Garching, Germany }
\date{\today}
\maketitle
\tightenlines
\begin{abstract}
Implications of field quantization on Ramsey interferometry are discussed and
general conditions for the occurrence of interference are obtained. Interferences
do not occur if the fields in two Ramsey zones have precise number of photons.
However in this  case it is shown how an analog of Hanbury-Brown Twiss photon-photon correlation
 interferometry can be used to
discern a variety of interference effects as the two independent Ramsey zones
get entangled by the passage of first atom. Interferences are restored by working with fields at a single photon
level. Generation of entangled states including states like $|2,0\rangle+e^{i\theta}|0,2\rangle$ and
$|\alpha,\beta\rangle+|-\alpha,-\beta\rangle$ is discussed.
\end{abstract}
\newpage

\section{Introduction}
The method of using spatially separated fields as proposed by Ramsey
 has been proved to be very useful for high resolution
work. It was originally proposed as a technique in the microwave domain
\cite{1} which
was then extended to studies in optical domain \cite{2}. More recently, this technique has
been used very successfully in the studies of quantum entanglement resulting
from the interaction of atoms with radiation in a high quality cavity. In this
context Ramsey technique enables one to study the entanglement in different
basis of states\cite{3}. Haroche and coworkers \cite{{3},{4},{5}} detected a variety of cavity quantum
electrodynamics effects \cite{berman} by monitoring changes in Ramsey fringe pattern when a 
cavity was placed in region between two Ramsey zones. Ramsey technique has also been suggested 
for the measurement
of phase diffusion in a micromaser \cite{6}. The existence of fringes in Ramsey
technique has been interpreted as due to quantum interferences in the transition
amplitudes and thus Ramsey technique is a way of doing atomic interferometry
\cite{7}. All these studies
consider the field in each Ramsey zone as a coherent field which is prescribed
and which does not evolve even though it is interacting with the atom.
Now that the interference effects at a single photon or few photon levels are
becoming quite common \cite{{9},{hariharan},{prl1},{prl2},{surendra}}, it is natural to enquire how the results of Ramsey
interferometry would be modified if the coherent field in each Ramsey zone is
replaced by a quantized field \cite{10}.

The outline of the paper is as follows. In Sec.II, we examine the theory of Ramsey interferometry with
quantized fields. In Sec.III, we bring out the role of the quantum statistics of the fields
in Ramsey interferometry and  examine the conditions on the fields so that interference
fringes are obtained. If the fields in the two Ramsey zones are in a state with
fixed number of photons, then interferences disappear \cite{scully}. Though fields with exactly
one photon do not yield interferences, a small coherent field even at a single
photon level can lead to a well-defined interference pattern. Further, we show photon-photon interaction 
can be mediated by a single atom passing through the two cavities. In Sec.IV, we compare this work to
our previous work\cite{10} and recent experiment of Bertet etal.\cite{5} on
complementarity and quantum eraser. In Sec.V, we
demonstrate an analog of Hanbury-BrownTwiss experiment in the context of Ramsey
interferometry with quantized fields by calculating atom-atom
correlations \cite{8} in
atoms passing through two successive Ramsey zones containing quantized field
\cite{prl2}. In the case 
of fixed number of photons in the cavities, interference does not occur in the excitation probability of a single
atom. We show how the interference can be observed by passing successively two atoms through the cavities and by
detecting both atoms in the excited states.
Finally, we consider the effect of decoherence in Ramsey interferometry.
We demonstrate the entanglement of the fields in the two Ramsey zones. Such
an entanglement results from the passage of an atom \cite{surendra}. We generate various entangled states 
by passing a single or two atoms through cavities and show  entanglement can be
transfered from fields to atoms. In Sec.VI, we consider the atom-cavity interaction in the dispersive limit 
and show the generation of  the entangled state of two coherent fields of the form
$|\alpha,\beta\rangle+|-\alpha,-\beta\rangle$. 
\section{Ramsey Interferometry in Quantized Fields}
        We consider a high quality cavity \cite{berman} as the Ramsey
zone of quantized field. If the number of photons in the cavity is large and
the field has a well-defined phase, then it would approach the classical Ramsey
interferometry. We thus consider the situation shown in the Fig.1. An atom with
two levels $|e\rangle$ and $|g\rangle$ interacts with two single mode cavities with identical frequencies. Let the
annihilation and creation operators in the i-th cavity be denoted by $a_{i}$ and
$a_{i}^{\dag}$, respectively. For the situation shown in the Fig.1, the 
Hamiltonian in the interaction picture is
\begin{eqnarray}
H_{1}&=&\hbar g_{1}(|e\rangle\langle g|a_{1}e^{i\Delta t}+a_{1}^{\dag}|g\rangle\langle e|
e^{-i\Delta t})~~~~~~~~~~~~~0<t\leq\tau_{1}~,\nonumber\\
H_{1}&=&0 \label{eq1}~~~~~~~~~~~~~~~~~~~~~~~~~~~~~~~~~~~~~~~~~~~~~~~~~~~~~
\tau_{1}<t\leq T+\tau_{1}~,\\
H_{1}&=&\hbar g_{2}(|e\rangle\langle g|a_{2}e^{i\Delta t}+a_{2}^{\dag}|g\rangle\langle e|
e^{-i\Delta t})~~~~~~T+\tau_{1}<t\leq T+\tau_{1}+\tau_{2}~. \nonumber
\end{eqnarray}
Here $\Delta=\omega_{0}-\omega_{1}$ and
 $g_{i}$ is the coupling constant of the atom with the vacuum in the i-th
cavity. Let us consider an initial state with atom in the lower state $|g\rangle$ and
the fields characterized by the state
$\sum_{n,\mu}F_{n,\mu}|n,\mu\rangle$. Here $|n\rangle
(|\mu\rangle)$ represents the Fock state in first(second) cavity. Let
$\phi_{e},\phi_{g}$ be the phase shifts in $|e\rangle$ and $|g\rangle$ which we
might introduce using same external perturbation between the cavities. Using the
interaction Hamiltonian $(\ref{eq1})$ the time evolution of the state can be calculated. The
state of the atom and cavity fields is found to be
\begin{eqnarray}
|\psi(\tau_{1}+T+\tau_{2})\rangle &=&\sum_{n,\mu} \left[F_{n,\mu}
C_{n-1}(\tau_{1})C_{\mu-1}(\tau_{2})
\exp \left(-i\Delta(\tau_{1}+\tau_{2})/2-i\phi_{g} \right)\right.\nonumber\\  
&+&\left.F_{n+1,\mu-1}S_{n}(\tau_{1})S_{\mu-1}(\tau_{2})\exp\left(-i\Delta(\tau_{1}+\tau_{2}+2T)/2
-i\phi_{e}\right)\right]|g,n,\mu\rangle\nonumber \\
&+&\sum_{n,\mu} \left[F_{n+1,\mu}
S_{n}^{*}(\tau_{1})C_{\mu}^{*}(\tau_{2})\exp\left(i\Delta(\tau_{1}+\tau_{2})/2-i\phi_{e}
\right)\right.\nonumber\\  
&~&+\left.F_{n,\mu+1}C_{n-1}(\tau_{1})S_{\mu}^{*}(\tau_{2})\exp\left(i\Delta(\tau_{1}+\tau_{2}+2T)/2
-i\phi_{g}\right)\right]|e,n,\mu\rangle~,
\label{eq2}
\end{eqnarray}
where
\begin{eqnarray}
C_{\alpha}(\tau)&=&\cos\left(\Omega_{\alpha}\tau/2\right)+\frac{i\Delta}{\Omega_{\alpha}}\sin\left(\Omega_{\alpha}\tau/2\right)\nonumber\\
S_{\alpha}(\tau)&=&
\frac{2ig_{\alpha}\sqrt{\alpha+1}}{\Omega_{\alpha}}\sin\left(\Omega_{\alpha}\tau/2\right),\nonumber\\
\Omega_{\alpha}&\equiv&\sqrt{(\Delta^{2}+4g_{\alpha}^{2}(\alpha+1)} ~~,
\label{eq3}
\end{eqnarray}
\hspace*{5cm} $\alpha=n,\mu$ and $g_{n}=g_{1}$, $g_{\mu}=g_{2}$.\\
The functions $C_{\alpha}$ and $S_{\alpha}$ describe the dynamics of the atom
interacting with a single mode cavity with initial state as a Fock state . Note
that $C_{\alpha}$($S_{\alpha}$) gives the probability amplitude of finding the
atom in the excited(ground) state given that it was in the excited state at time
$t=0$.
 
The structure of the state clearly suggests the elementary process
responsible for the final state. A given final state is reached in two different
ways. Consider a measurement in which the outgoing
atom is found in the excited state. The probability of excitation $P_{eg}$ 
defined by
\begin{equation}
P_{eg}=Tr_{field}\langle e|\psi(T+\tau_{1}+\tau_{2})\rangle\langle
\psi(T+\tau_{1}+\tau_{2})|e\rangle
\label{eq4}
\end{equation}
 can be calculated using Eq.$(\ref{eq2})$. We find the result 
\begin{equation}
P_{eg}= \sum_{n,\mu}|F_{n+1,\mu}
X_{n+1,\mu}+e^{i\Delta T+i\phi}
F_{n,\mu+1}Y_{n,\mu+1}|^{2}~,
\label{eq5}
\end{equation}
where
\begin{eqnarray}
\phi &=&\phi_{e}-\phi_{g}\nonumber\\
X_{n+1,\mu}&=&S_{n}^{*}(\tau_{1})C_{\mu}^{*}(\tau_{2})\nonumber\\
Y_{n,\mu+1}&=&C_{n-1}(\tau_{1})S_{\mu}^{*}(\tau_{2})~.
\label{eq6}
\end{eqnarray}

A similar result is obtained for $P_{ge}$ i.e. the probability of finding the
atom in the ground state 
if initially the atom is in the excited state,
\begin{equation}
P_{ge}= \sum_{n,\mu}|F_{n-1,\mu}
X^{*}_{n,\mu-1}+e^{-i\Delta T+i\phi}
F_{n,\mu-1}Y^{*}_{n+1,\mu}|^{2}~.
\label{eq7}
\end{equation}
In particular if each cavity is in vacuum state and $g_{1}=g_{2}, \tau_{1}=\tau_{2}$ then we get much simpler result,
\begin{equation}
P_{ge}=\frac{4g^{2}}{\Omega_{0}^{2}}\sin^{2}\frac{\Omega_{0}\tau_{1}}{2}\left(2-\frac{4g^{2}}{\Omega_{0}^{2}}
\sin^{2}\frac{\Omega_{0}\tau_{1}}{2}\right),
\label{eq8}
\end{equation}
which exhibits no interferences as it is phase independent.

The results $(\ref{eq5})$ and $(\ref{eq7})$ are important for understanding Ramsey interferometry with quantized
fields. These  give rise to a number of important consequences as far as the fundamentals of 
atom-field interaction are
concerned. For classical fields result$(\ref{eq7})$ can be  modified, as probability amplitude functions $F_{n,\mu}$ is peaked
around average number of photons $\bar n$ and $\bar\mu$. So in the summation we can replace
\begin{eqnarray}
X_{n,\mu-1} \rightarrow X_{\bar n,\bar\mu},\nonumber\\
Y_{n+1,\mu} \rightarrow Y_{\bar n,\bar\mu}.
\end{eqnarray}
Further for large $n$ and $\mu$, we replace,
\begin{eqnarray}
F_{n-1,\mu+1} \rightarrow F_{n,\mu}\nonumber\\
F_{n,\mu-1} \rightarrow F_{n,\mu}.
\end{eqnarray}
For normalised photon probability amplitude functions $F_{n,\mu}$ Eq.$(\ref{eq7})$ reduces to,
\begin{equation}
P_{ge}= |X^{*}_{\bar n,\bar\mu}+e^{-i\Delta T+i\phi}Y^{*}_{\bar n,\bar\mu}|^{2}~.
\label{eq7new}
\end{equation}
Equation $(\ref{eq7new})$ is the result for classical fields.
\section{Dependence of the fringes on quantum statistics of the fields}
 We now examine the consequences of the quantized nature of the field, and in particular investigate when the
 interferences are most pronounced. From the result$(\ref{eq5})$, we see there are two paths which contribute 
 to the amplitude for detecting
the atom in excited state
\begin{eqnarray}
|g,n,\mu\rangle&\rightarrow&|e,n-1,\mu\rangle\rightarrow|e,n-1,\mu\rangle~,
\nonumber\\
|g,n,\mu\rangle&\rightarrow&|g,n,\mu\rangle\rightarrow|e,n,\mu-1\rangle~.
\label{eq7b}
\end{eqnarray}
The interference between these two paths  depends on the nature of the photon
statistics i.e. on the functions $F_{n,\mu}$. Clearly if the field in each
cavity is in a Fock state $|n_{0},\mu_{0}\rangle$
\begin{equation}
F_{n,\mu}=\delta_{n,n_{0}}
\delta_{\mu,\mu_{0}}~,
\label{eq8b}
\end{equation}
then the interference terms in $(\ref{eq5})$ drop out and the two paths
$(\ref{eq7})$ become independent. This happens even for Fock states with large number of
photons. Interferences are obtained as long as the photon statistics is such that
\begin{equation}
\sum_{n,\mu}F_{n+1,\mu}F^{*}_{n,\mu+1}X_{n+1,\mu}Y^{*}_{n,\mu+1}\neq0~.
\label{eq9}
\end{equation}
In order to understand the meaning of Eq.(9), consider a situation where detuning $\Delta$ can
be ignored while considering evolution in Ramsey zone i.e. in each cavity. The condition (9) can be
reduced to a very interesting form
\begin{equation}
\langle a_{1}^{\dag}\frac{1}{\sqrt{a_{1}a_{1}^{\dag}}}\sin(g_{1}\tau_{1}\sqrt{a_{1}a_{1}^{\dag}}
)\cos(g_{1}\tau_{1}\sqrt{a_{1}^{\dag}a_{1}})\cos(g_{2}\tau_{2}\sqrt{a_{2}
a_{2}^{\dag}})\frac{1}{\sqrt{a_{2}a_{2}^{\dag}}}\sin(g_{2}\tau_{2}\sqrt{a_{2}a_{2}^{\dag}})
a_{2}\rangle \neq0~,
\label{eq10}
\end{equation}
 which for small interaction times reduces to
\begin{equation}
\langle a_{1}^{\dag}a_{2}\rangle \neq0~.
\label{eq11}
\end{equation}
Thus the nature of interference depends on the quantum statistics of the fields
in the two Ramsey zones.
The conditions $(\ref{eq10})$ and $(\ref{eq11})$ imply that if the cavities are independent, then the
field in each cavity must have a well defined phase for interference to occur. The interference would 
also not occur if one cavity has a definite
number of photons and the other has a field in coherent state. However,
interference is obtained if  fields in the two cavities are entangled
even though  the field in each cavity does not have a well-defined phase. In Fig.\ref{fig5}, results for classical
as well as quantized fields are plotted when each Ramsey zone has a coherent field with average number of photons
 $(|\alpha|^{2}=5)$.
Interference fringes for classical fields show higher visibility than in the case of quantized fields. 
 
\subsection{Ramsey fringes with fields at single photon level}
Having shown that Ramsey fringes vanish if each cavity contains one photon, the
next question arises what happens if the field in each cavity is at single
photon level\cite{hariharan}. For this purpose, we consider a case where each cavity is pumped by a weak coherent state so that
the initial state of the cavities is 
\begin{equation}
|\psi_{cavities}\rangle\cong \frac{1}{(1+{|\alpha|}^{2})}(|0\rangle+\alpha|1\rangle)
(|0\rangle+\alpha e^{i\theta}|1\rangle).
\label{eq12}
\end{equation}
In this case, the result $(\ref{eq5})$ leads to 
\begin{eqnarray}
P_{eg}&=&\frac{4|\alpha|^{2}}{(1+|\alpha|^{2})^{2}}\left|\frac{g_{1}}{\Omega_{0}}\sin(\Omega_{0}\tau_{1}/2)
\left(\cos(\Omega_{0}\tau_{2}/2)-\frac{i\Delta}{\Omega_{0}}
\sin(\Omega_{0}\tau_{2}/2)\right) \right. \nonumber\\
 &+&\left.\frac{g_{2}}{\Omega_{0}}\sin(\Omega_{0}\tau_{2}/2)
\exp[i\{\Delta (T+\tau_{1}/2)+\theta+\phi\}]\right| ^{2}+O(|\alpha|^{4})~,
\label{eq13}
\end{eqnarray}
which for  $\Delta=0$ and $g_{1}\tau_{1}=\sqrt{2}g_{2}\tau_{2}=\pi/2$ reduces to
\begin{equation}
P_{eg}=\frac{|\alpha|^{2}}{(1+|\alpha|^{2})^{2}}\left(1+\sin(\pi/\sqrt{2})\cos(\theta+\phi)\right).
\label{eq14}
\end{equation}
This leads to high visibility for the fringes(about$80\%$).
It is clear that the interference in $(\ref{eq13})$ arises from the cross terms in $(\ref{eq12})$  
as such cross terms lead to the same final state via two different pathways 
\begin{eqnarray}
|g,1,0\rangle\rightarrow|e,0,0\rangle\rightarrow|e,0,0\rangle~,
\nonumber\\
|g,0,1\rangle\rightarrow|g,0,1\rangle\rightarrow|e,0,0\rangle.
\end{eqnarray}
The other terms in $(\ref{eq12})$ do not result in interference as $|1,1\rangle$
leads to different final states  and $|0,0\rangle$ can not produce
excitation. It should be borne in mind that $(\ref{eq13})$ is different from the
result obtained for classical fields.
For classical fields on resonance $\Delta=0$, $P_{eg}$ is given by,
\begin{eqnarray}
P_{eg}=\left|\sin(\Omega_{R}\tau_{1}/2)\cos(\Omega_{R}\tau_{2}/2)
 +\sin(\Omega_{R}\tau_{2}/2)\cos(\Omega_{R}\tau_{1}/2)
\exp\{i(\theta+\phi)\}\right| ^{2}~.
\label{eq3n}
\end{eqnarray}
Here $\Omega_{R}$ is Rabi frequency, $\theta$ is phase difference between the two fields . To compare the result
for classical fields with the  result for quantized fields, one can use approximation $\Omega_{R}=2g|\alpha|$~.
For comparing Eq.$(\ref{eq14})$ to the result for classical fields  we  put 
$g\tau_{1}=g\sqrt{2}\tau_{2}=\pi/2$ in $(\ref{eq3n})$. The result $(\ref{eq3n})$ 
 reduces to
\begin{equation}
P_{eg}=\left|\sin\left(\frac{\pi|\alpha|}{2}\right)\cos\left(\frac{\pi|\alpha|}
{2\sqrt{2}}\right)
+\sin\left(\frac{\pi|\alpha|}{2\sqrt{2}}\right)\cos\left(\frac{\pi|\alpha|}{2}
 \right)
\exp\{i(\theta+\phi)\}\right|^{2}~.
\label{eq3new}
\end{equation}
For small $|\alpha|$ the above result takes the form,
\begin{equation}
P_{eg}=\frac{|\alpha|^{2}\pi^{2}}{4}\left(\frac{3}{2}+\sqrt{2}\cos(\theta+\phi)\right).
\label{eq14b}
\end{equation}
The Eq.$(\ref{eq14})$ shows nearly $80\%$ visibility while the result $(\ref{eq14b})$ gives nearly $95\%$
visibility. 
For weak classical fields Ramsey interference pattern shows higher visibility than the results
derived with quantized fields. The result for classical fields can have $100\%$
visibility if both cavities have the same coherent fields and $\Omega_{R}\tau_{1}=\Omega_{R}\tau_{2}$.
\subsection{Ramsey fringes with fields in nonclassical state}
Recently an arbitrary superposition of
$|0\rangle$ and $|1\rangle$ states $\frac{1}{\sqrt{1+|\alpha|^{2}}}(|0\rangle+\alpha|1\rangle)$ has been realised 
\cite{{prl1},{prl2}}, here the
parameter $\alpha$ need not be small. Such a state is highly nonclassical and is quite distinct from a coherent
state with very small excitation. This nonclassical state also has the
important characteristics that the average value of the field is nonzero and thus the off-diagonal elements of the
density matrix are nonzero. For such a nonclassical state and for $\Delta=0$, the Ramsey fringes are given by,
\begin{eqnarray}
P_{eg}&=&\frac{|\alpha|^{2}}{(1+|\alpha|^{2})^{2}}\left|\sin(g_{1}\tau_{1})\cos(g_{2}\tau_{2})
 +\sin(g_{2}\tau_{2})
\exp\{i\phi\}\right| ^{2}\nonumber\\
&+&\frac{|\alpha|^{4}}{(1+|\alpha|^{2})^{2}}\left\{\sin^{2}(g_{1}\tau_{1})\cos^{2}(g_{2}\sqrt{2}\tau_{2})
+\cos^{2}(g_{1}\tau_{1})\sin^{2}(g_{2}\tau_{2})\right\}
\label{eq14new}
\end{eqnarray}
For $g_{1}\tau_{1}=g_{2}\sqrt{2}\tau_{2}=\pi/2$, it reduces to the previously
derived result $(\ref{eq14})$. Remarkably the visibility from the result$(\ref{eq14new})$ does not depend on $\alpha$ 
which is in contrast to the result
$(\ref{eq3new})$ . We show in the Fig(\ref{graph1}) a comparison of the visibility in the case of coherent field and a
nonclassical field. It may be noted that for the state $|0\rangle+\alpha|1\rangle$, the P-distribution is quite
singular.   
\subsection{Photon-photon interaction mediated by a single atom and quantum entanglement of two cavities}
It is well known in nonlinear optics in a macroscopic system that the fields
effectively interact and one knows many examples of three wave and four wave
interactions in a medium. Such interactions are significant at macroscopic densities of
atoms. In this section, we demonstrate a rather remarkable result that a single
atom in a high quality cavity can produce photon-photon interaction. For this
purpose consider an atom in the ground state passing through the two cavity
system. We calculate the state of the two cavity system subject to the condition
that the atom at the output is detected in the ground state. Such a conditional
field state is found to be,
\begin{eqnarray}
|\psi_{c,g}\rangle&=&\langle g|\psi(T+\tau_{1}+\tau_{2})\rangle~,\nonumber\\
&=& C_{n-1}(\tau_{1})C_{\mu-1}(\tau_{2})\exp
\left(-i\Delta(\tau_{1}+\tau_{2})/2-i\phi_{g} \right)|n,\mu\rangle\nonumber\\  
&+&S_{n-1}(\tau_{1})S_{\mu}(\tau_{2})\exp\left(-i\Delta(\tau_{1}+\tau_{2}+2T)/2-i\phi_{e}\right)
|n-1,\mu+1\rangle~.
\end{eqnarray}
This involves a linear combination of states $|n,\mu\rangle$ and 
$|n-1,\mu+1\rangle$ leading to the entanglement of two cavities. In addition, the
passage of one atom transfers one photon from the first cavity to the second
cavity. The transfer from one cavity to the other will be complete if $C_{n-1}(\tau_{1})=C_{\mu-1}(\tau_{2})=0$.
Other entangled states are possible, for example, if the atom was initially in the ground state and if it was detected in the
excited state, then the conditional state of the cavities is,
\begin{eqnarray}
|\psi_{c,e}\rangle&=&\langle e|\psi(T+\tau_{1}+\tau_{2})\rangle~,\nonumber\\
&=& S_{n-1}^{*}(\tau_{1})C_{\mu}^{*}(\tau_{2})\exp
\left(i\Delta(\tau_{1}+\tau_{2})/2-i\phi_{e} \right)|n-1,\mu\rangle\nonumber\\  
&+&C_{n-1}(\tau_{1})S_{\mu-1}^{*}(\tau_{2})\exp\left(i\Delta(\tau_{1}+\tau_{2}+2T)/2-i\phi_{g}\right)
|n,\mu-1\rangle~.
\label{new}
\end{eqnarray} 
\section{Relation to the earlier works}
Before we proceed further we
compare the above with our previous work \cite{10} and a recent experiment of
Bertet etal \cite{5}. In our
earlier work we used a perturbative approach valid for short interaction times.For most part we also  assumed
that the two regions of Ramsey interaction were identical in all respects. Thus the atoms interacted with
the same quantized field in each region. In such a situation, the interference
always occurs as the condition
$(\ref{eq11})$ is replaced by $\langle a^{\dag}a\rangle \neq0$ . Thus in our previous investigation, the
effects of quantized field did not have any effect on Ramsey fringes. However, the statistics of quantized
field affects the quantum noise in the measured signal. In the current paper, we deal with the case of
strong atom-cavity interaction; besides the two cavities do not necessarily  have identical fields.  Thus the quantum
statistics is very relevant even for Ramsey fringes in the excitation probability as is borne out by the examples given
in  the Sect.III.

The recent experiment of Bertet etal \cite{5} on complementarity and quantum eraser is a realization of Ramsey
interferometry with quantized fields. These authors create two Ramsey zones in the same cavity Fig.\ref{fig2},
 by having a region in which atom undergoes a phase shift. The fringes are restored
for this situation as the two Ramsey zones are located in the same cavity,
 a case for which condition 
$(\ref{eq11})$ is always fulfilled.  Let us assume
that the ground state of atom $|g\rangle$ undergoes phase change
after time $\tau_{1}$ by some internal perturbation but excited state $|e\rangle$ remains same and total 
interaction time for an atom is $\tau_{1}+\tau_{2}$ . The field inside the cavity is in state 
$\sum_{n}F_{n}|n\rangle$.
A single atom initially in the excited $|e\rangle$ passes through the cavity and
detected in the ground state $|g\rangle$. 
The probability $P_{ge}$ is given by, 
\begin{equation}
P_{ge}=\sum_{n}\left|F_{n}S_{n}(\tau_{1})C_{n}(\tau_{2})\exp(-i\phi)+F_{n}C^{*}_{n}(\tau_{1})S_{n}(\tau_{2})\right|^{2}.
\end{equation}
If the cavity has coherent field with average number of photons equal to $\bar n$ and $\tau_{1}=\tau_{2}$, 
$\Delta=0$, then the above expression leads to the simple result,
\begin{equation}
P_{ge}=\sum_{n}\frac{\bar n^{n}e^{-\bar n}}{n!}\sin^{2}(2g\sqrt{n+1}\tau_{1})\cos^{2}(\phi/2).
\end{equation}
The phase factor appears as an overall multiplication factor. Thus interferences always occur even if the field in the
cavity is in Fock state. This situation is quite different from the one involving two cavities $(\ref{eq7})$.

\section{Two Atom Interferometry}
In Sect.III, we considered the possibility of producing entanglement between the two cavities by conditional
detection of the atomic state. We next examine how such entanglement $(\ref{new})$
can be detected. From our previous discussion leading to
$(\ref{eq9})$, $(\ref{eq10})$ it is clear that if we send a second atom and measure its excitation
probability then such a probability would exhibit interference fringes.

At the outset, we mention that a large body of literature exists on atom-atom correlations and entanglement, when
such atoms fly through a high quality cavity. The cavity could be a closed one as in Garching experiments \cite{8}
or an open one as in Paris experiments \cite{3}. For a second atom coming in the ground state $|g\rangle$ and detected in the excited state
following are the possible pathways,
\begin{eqnarray}
|n-1,\mu\rangle|g\rangle&\rightarrow&|n-1,\mu\rangle|g\rangle\rightarrow|n-1,\mu-1\rangle|e\rangle~,
\nonumber\\
|n,\mu-1\rangle|g\rangle&\rightarrow&|n-1,\mu-1\rangle|e\rangle\rightarrow|n-1,\mu-1\rangle|e\rangle~,
\nonumber\\
|n-1,\mu\rangle|g\rangle&\rightarrow&|n-2,\mu\rangle|e\rangle\rightarrow|n-2,\mu\rangle|e\rangle~,
\nonumber\\
|n,\mu-1\rangle|g\rangle&\rightarrow&|n,\mu-1\rangle|g\rangle\rightarrow|n,\mu-2\rangle|e\rangle~.
\nonumber\\
\end{eqnarray}
 In summary, the system as a whole  starting with an initial state $|n,\mu,g_{1},g_{2}\rangle$ 
 has two different pathways leading to the detection of the atom-cavity system in 
 state $|n-1,\mu-1,e_{1},e_{2}\rangle$.
\begin{eqnarray}
|n,\mu,g_{1},g_{2}\rangle&\rightarrow&|n-1,\mu,e_{1},g_{2}\rangle\rightarrow|n-1,\mu-1,e_{1},e_{2}\rangle~,
\nonumber\\
|n,\mu,g_{1},g_{2}\rangle&\rightarrow&|n,\mu-1,e_{1},g_{2}\rangle\rightarrow|n-1,\mu-1,e_{1},e_{2}\rangle~.
\nonumber\\
\end{eqnarray} 
   The joint probability
of detecting both atoms in the excited state $P_{g_{1}g_{2}}^{e_{1}e_{2}}$ can be used for doing atomic
interferometry even if each cavity is in Fock state. This is reminiscent of 
 photon-photon correlation measurements with light produced in the process of
 down conversion. Mandel and coworkers \cite{mandel} carried out a series of measurements with photons from
 a down converted source where they reported no interferences in the measurement
 of mean intensities whereas photon-photon correlation exhibited a variety of
 interference phenomena. In the context of Ramsey interferometry with quantized
fields we suggest a measurement of the atom-atom correlation. An explicit form of the joint detection
probability can be obtained following Jaynes-Cummings dynamics. 
A long calculation leads to the
following expression for the joint probability if the initial state of the cavities is $|n,\mu\rangle$
\begin{eqnarray}
P_{g_{1}g_{2}}^{e_{1}e_{2}}=\left|S_{n-1}(\tau_{1})S_{n-2}(\tau_{1}^{'})C_{\mu}^{*}(\tau_{2})C_{\mu}^{*}
(\tau_{2}^{'})\right|^{2}
+\left|C_{n-1}(\tau_{1})C_{n-1}(\tau_{1}^{'})S_{\mu-1}(\tau_{2})S_{\mu-2}(\tau_{2}^{'})\right|^{2}\nonumber\\
+\left|S_{n-1}(\tau_{1}^{'})C_{n-1}(\tau_{1})S_{\mu-1}(\tau_{2})C_{\mu-1}^{*}(\tau_{2}^{'})
+S_{n-1}(\tau_{1})C_{n-2}(\tau_{1}^{'})S_{\mu-1}(\tau_{2}^{'})C_{\mu}^{*}(\tau_{2})\right.\nonumber\\
\left.\exp[i(\Delta(T^{'}-T)+\phi^{'}-\phi)]\right|^{2}~.
\label{eq15}
\end{eqnarray}
Here we allow the possibility of different interaction times and phases(denoted by dash) for the second atom.
In the special case when $\Delta=0,g_{1}=g_{2}, \tau_{1}=\tau_{2}=\tau_{1}^{'}
=\tau_{2}^{'}$ and $g\tau=\pi/4$ and when initially cavities are in state $|1,1\rangle$, the joint detection
 probability has the form
\begin{eqnarray}
P_{g_{1}g_{2}}^{e_{1}e_{2}}&=&\frac{1}{16}+\frac{1}{4}\cos^{2}(\pi/2\sqrt{2})+\frac{1}{4}\cos(\pi/2\sqrt{2})
\cos(\phi^{'}-\phi)~\nonumber\\
&=&0.1118+0.1110\cos(\phi^{'}-\phi).
\label{eq16}  
\end{eqnarray}
Interference fringes with almost $100\%$ visibility are obtained. Thus two atom interferometry could produce
perfect visibility in the situations where single atom interferometry exhibits no interferences. Other joint detection probabilities like 
finding one atom in the excited state and the other in the ground state also display interference fringes.
An interesting situation also corresponds to sending both atoms in the excited state and measuring the final states
of the two atoms. In the case when initially cavities are in the state
$|0,0\rangle$ and $\Delta=0$, the expression for the probability of detecting both the atoms
in their ground states has the form
\begin{eqnarray}
P_{e_{1}e_{2}}^{g_{1}g_{2}}&=&\sin^{2}(g_{1}\tau_{1})\sin^{2}(g_{1}\sqrt{2}\tau_{1}^{'})+
\sin^{2}(g_{2}\tau_{2})\sin^{2}(g_{2}\sqrt{2}\tau_{2}^{'})\cos^{2}(g_{1}\tau_{1})\cos^{2}(g_{1}\tau_{1}^{'})
\nonumber\\
&+&\left|\cos(g_{1}\tau_{1})\sin(g_{1}\tau_{1}^{'})\sin(g_{2}\tau_{2})\cos(g_{2}\tau_{2}^{'})+
\sin(g_{1}\tau_{1})\cos(g_{1}\sqrt{2}\tau_{1}^{'})\sin(g_{2}\tau_{2}^{'})
e^{i(\phi-\phi^{'})}\right|^{2}.
\label{eq20}
\end{eqnarray}
Consider the case when  $g_{1}\sqrt{2}\tau_{1}^{'}=g_{2}\sqrt{2}\tau_{2}^{'}=\pi$ and  $g_{1}\tau_{1}=g_{2}\tau_{2}
=\pi/4$.
The probability of detecting both atoms in the ground states is given by
\begin{eqnarray}
P_{e_{1}e_{2}}^{g_{1}g_{2}}&=&\frac{1}{2}\left|\frac{1}{\sqrt{2}}\sin{\frac{\pi}{\sqrt{2}}}\cos{\frac{\pi}{\sqrt{2}}}-
\sin{\frac{\pi}{\sqrt{2}}}\exp[i(\phi-\phi^{'})]\right|^{2},\nonumber\\
&=&0.4327+0.3835\cos(\phi-\phi^{'}).
\label{eq17}  
\end{eqnarray}
The visibility of fringes in two atom interferometry is quite significant. We next show how two atom interferometry
can be used to produce a variety of entangled states.
\subsection{Preparation of the entangled state $\alpha|2,0\rangle+\beta|0,2\rangle$}
Consider the situation when two identical atoms are coming in their excited
states and each cavity is in vacuum state. The mode of each cavity is in 
resonance with the atomic transition frequency. If after passing through the cavities both atoms are detected in 
their ground states, the state of the field inside the cavities is given by
\begin{eqnarray}
|\Phi(\tau_{1}+T+\tau_{2})\rangle&=&\sin(g_{1}\tau_{1})\sin(g_{1}\sqrt{2}\tau_{1}^{'})\exp\{-i(\phi_{g}+\phi^{'}_{g})\}
|2,0\rangle \nonumber\\
&+&\cos(g_{1}\tau_{1})\cos(g_{1}\tau_{1}^{'})\sin(g_{2}\tau_{2})\sin(g_{2}\sqrt{2}\tau_{2}^{'})
\exp\{-i(\phi_{e}+\phi_{e}^{'})\}|0,2\rangle\nonumber\\
&+&\left[\cos(g_{1}\tau_{1})\sin(g_{1}\tau_{1}^{'})\sin(g_{2}\tau_{2})\cos(g_{2}\tau_{2}^{'})
\exp\{-i(\phi_{e}+\phi_{g}^{'})\}\right.\nonumber\\
&+&\left.\sin(g_{1}\tau_{1})\cos(g_{1}\sqrt{2}\tau_{1}^{'})\sin(g_{2}\tau_{2}^{'})
\exp\{-i(\phi_{g}+\phi_{e}^{'})\}\right]|1,1\rangle ~.
\label{eqn21}
\end{eqnarray}
The $|1,1\rangle$ component drops out for $g_{1}\tau_{1}^{'}=g_{2}\tau_{2}^{'}=\pi$ and the cavities will be in the entangled state
\begin{eqnarray}
|\Phi(\tau_{1}+T+\tau_{2})\rangle&=&\sin(g_{1}\tau_{1})\sin(\pi\sqrt{2})\exp[-i(\phi_{g}+\phi^{'}_{g})]
|2,0\rangle \nonumber\\
&-&\cos(g_{1}\tau_{1})\sin(g_{2}\tau_{2})\sin(\pi\sqrt{2})
\exp[-i(\phi_{e}+\phi_{e}^{'})]|0,2\rangle.
\label{eqn22}
\end{eqnarray}
This entangled state is very interesting, we can change the degree of
entanglement by changing the value of $g_{1}\tau_{1}$ and
$g_{2}\tau_{2}$. The state will be maximally entangled for $g_{1}\tau_{1}=\pi/4$ and
$g_{2}\tau_{2}=\pi/2$. This can be seen as first atom comes in excited state
$|e\rangle$ interacts with the first cavity for a time such that
$g_{1}\tau_{1}=\pi/4$. The interaction is like an interaction with a $\pi/2$
pulse the state of the system evolves into
\begin{equation}
|e,0,0\rangle\rightarrow \frac{1}{\sqrt{2}}(|e,0,0\rangle-i|g,1,0\rangle).
\end{equation}
In the second cavity the interaction is a $\pi$ pulse interaction and then the
state of the total system becomes,
\begin{equation}
\frac{1}{\sqrt{2}}(|e,0,0\rangle-i|g,1,0\rangle)\rightarrow-\frac{i}{\sqrt{2}}(|g,1,0\rangle+|g,0,1\rangle).
\end{equation}
Thus after passing the first atom the state of the fields in the cavities is,
\begin{equation}
|\psi_{1}\rangle=-\frac{i}{\sqrt{2}}(|1,0\rangle+|0,1\rangle).
\label{field}
\end{equation}
The second atom comes in the excited state $|e\rangle$  interacts with the
fields inside the
cavities for times $\tau_{1}^{'}$ and $\tau_{2}^{'}$ such that
$g_{1}\tau_{1}^{'}=g_{2}\tau_{2}^{'}=\pi$  and after passing through the
cavities the atom is detected in the
ground state $|g\rangle$, so the atom can follow the two paths. The first path is,
\begin{eqnarray}
|e\rangle|1,0\rangle\rightarrow \cos(\pi\sqrt{2})|e,1,0\rangle-i\sin(\pi\sqrt{2})|g,2,0\rangle
\rightarrow -i\sin(\pi\sqrt{2})|g,2,0\rangle-\cos(\pi\sqrt{2})|e,1,0\rangle,
\end{eqnarray}
The second path is,
\begin{equation}
|e\rangle|0,1\rangle\rightarrow -|e,0,1\rangle\rightarrow i\sin(\pi\sqrt{2})|g,0,2\rangle-\cos(\pi\sqrt{2})|e,0,1\rangle.
\end{equation}
 Thus passing both the atoms initially in the excited states and subsequently detecting them in
their ground states, we obtain a
maximally entangled state
\begin{equation}
|\psi_{2}\rangle= \frac{1}{\sqrt{2}}\sin(\pi\sqrt{2})\left(\exp[-i(\phi_{g}+\phi^{'}_{g})]|2,0\rangle
 -\exp[-i(\phi_{e}+\phi_{e}^{'})]|0,2\rangle\right), 
\label{ent}
\end{equation}
of the fields inside the cavities.
The phase terms in $(\ref{ent})$ come from the phase change in the region between the cavities.
Now if we pass another atom initially in the excited state $|e\rangle$ through the cavities
having field in state $(\ref{ent})$ and the atom is detected in its ground state $|g\rangle$ after passing through the
cavities, entangled state of three photons is generated. The degree of entanglement is
controlled by the selection of interaction times in the cavities. For a special case a three photons
maximally entangled state, 
\begin{equation}
|\psi_{3}\rangle=-\frac{i}{\sqrt{2}}\sin(\pi\sqrt{2})\sin(\pi\sqrt{3})
\left[e^{-i(\phi_{g}+\phi_{g}^{'}+\phi_{g}^{''})}|3,0\rangle 
+e^{-i(\phi_{e}+\phi_{e}^{'}+\phi_{e}^{''})}|0,3\rangle\right],
\label{eq40}
\end{equation}
is generated if we choose interaction times $\tau_{1}^{''}$ and $\tau_{2}^{''}$ for third atom
such that $g_{1}\tau_{1}^{''}=g_{2}\tau_{2}^{''}=\pi$.

We note that in any interference experiment involving superposition of quantum states, the decoherence of 
the state is an important issue. The well defined interferences occur provided that experimental time scales are much 
smaller compared to the decoherence time. Thus the preparation of the entangled state$(\ref{ent})$ would be
sensitive to the decoherence of the state$(\ref{field})$ in the interval $\tau_{0}$
between the detection of the first atom and the arrival of the second atom. Such
issues have been experimentally studied by Brune etal and Raimond etal\cite{{brune}}.
The decoherence of the state$(\ref{field})$ can be studied by solving the master
equation describing the decay of field in each cavity,
\begin{equation}
\dot{\rho}\equiv -\sum_{i=1}^{2} \kappa_{i}\left(a_{i}^{\dag}a_{i}\rho -2a_{i}\rho
a_{i}^{\dag}+\rho a_{i}^{\dag}a_{i}\right),
\label{meq}
\end{equation}
where $2\kappa_{i}$ gives the rate of loss of photons from the ith cavity.
Under the initial condition$(\ref{field})$ a long calculation shows that,
\begin{eqnarray}
|\psi_{1}\rangle\langle\psi_{1}|&~&\rightarrow|\psi_{1t}\rangle\langle\psi_{1t}|+
\left(1-\frac{1}{2}(e^{-2\kappa_{1}t}+e^{-2\kappa_{2}t})\right)|0,0\rangle\langle0,0|,\nonumber\\
|\psi_{1t}\rangle&=&-\frac{i}{\sqrt{2}}\left(|1,0\rangle e^{-\kappa_{1}t}+|0,1\rangle
e^{-\kappa_{2}t}\right).
\label{decoh}
\end{eqnarray}
Thus decoherence will not be a serious issue as long as $\kappa_{i}\tau_{0}<<1$. 
\subsection{Entanglement Transfer from Fields to Atoms}
Here we show how entanglement of fields is transfered to the atoms. For this purpose consider
the fields inside the cavities are in an entangled state,
\begin{equation}
|\psi_{cf}\rangle=\alpha|0,1\rangle+\beta|1,0\rangle.
\end{equation}
 An atom initially in the ground state $|g\rangle$ is passed through the cavities and the fields inside
the cavities are in resonance with atomic transition frequency, then the state of
the cavity-atom system is,
\begin{eqnarray}
|\psi_{4}\rangle&=&\left\{\alpha\cos(g_{2}\tau_{2})e^{-i\phi_{g}}
-\beta\sin(g_{1}\tau_{1})\sin(g_{2}\tau_{2})e^{-i\phi_{e}}\right\}|g,0,1\rangle
+\beta\cos(g_{1}\tau_{1})e^{-i\phi_{e}}|g,1,0\rangle\nonumber\\
&-&i\left(\alpha\sin(g_{2}\tau_{2})e^{-i\phi_{g}}+
\beta\sin(g_{1}\tau_{1})\cos(g_{2}\tau_{2})e^{-i\phi_{e}}\right)|e,0,0\rangle.
\label{eq41}
\end{eqnarray}
If another atom coming in the ground state $|g\rangle$, interacts with the fields in
both the cavities for the times $\tau_{1}^{'}$ and $\tau_{2}^{'}$ such that
$g_{1}\tau_{1}^{'}=g_{2}\tau_{2}^{'}=\pi/2$, then the state of the cavity-atom system is,
\begin{eqnarray}
|\psi_{5}\rangle&=&-i\left(\alpha\sin(g_{2}\tau_{2})e^{-i(\phi_{g}+\phi_{g}^{'})}+
\beta\sin(g_{1}\tau_{1})\cos(g_{2}\tau_{2})e^{-i(\phi_{e}+\phi_{g}^{'})}\right)|e,g\rangle|0,0\rangle
\nonumber\\
&-&i\left(\alpha\cos(g_{2}\tau_{2})e^{-i(\phi_{g}+\phi_{g}^{'})}+
\beta\sin(g_{1}\tau_{1})\sin(g_{2}\tau_{2})e^{-i(\phi_{e}+\phi_{g}^{'})}\right)|g,e\rangle|0,0\rangle
\nonumber\\
&-&\beta\cos(g_{1}\tau_{1})e^{-i(\phi_{e}+\phi_{g}^{'})}|g,g\rangle|1,0\rangle.
\label{eq42}
\end{eqnarray}
If we choose the interaction time for first atom in first cavity such that
$g_{1}\tau_{1}=\pi/2$ the state $(\ref{eq42})$
becomes,
\begin{eqnarray}
|\psi_{6}\rangle&=&-i\left(\alpha\sin(g_{2}\tau_{2})e^{-i(\phi_{g}+\phi_{g}^{'})}+
\beta\cos(g_{2}\tau_{2})e^{-i(\phi_{e}+\phi_{g}^{'})}\right)|e,g\rangle|0,0\rangle
\nonumber\\
&-&i\left(\alpha\cos(g_{2}\tau_{2})e^{-i(\phi_{g}+\phi_{g}^{'})}+
\beta\sin(g_{2}\tau_{2})e^{-i(\phi_{e}+\phi_{g}^{'})}\right)|g,e\rangle|0,0\rangle.
\label{eq43}
\end{eqnarray}
The state $(\ref{eq43})$ shows that the atoms are now in entangled state and fields are in
independent states so the entanglement of fields has been transfered to the atoms. 
\section{creation of entangled states of coherent states}
Here we discuss the case when atom field coupling in the cavities is dispersive
in nature.  This is the case when atom-field detuning is very large  compared to 
coupling so that fields inside the cavities can produce phase
changes only\cite{brune} without any photon
absorption or emission. In such a case, a simple perturbative analysis shows that the state
$|e,n\rangle (|g,n\rangle)$ experiences the energy shift $\frac{\hbar
g^{2}(n+1)}{\Delta} (-\frac{\hbar g^{2}n}{\Delta})$. The effective Hamiltonian of 
atom-cavity system can be written as
\begin{eqnarray}
H&=&\sum_{n,\mu}\hbar\left[(n+\mu)\omega
+\frac{\omega_{0}}{2}+\frac{g^{2}(n+1)}{\Delta}\right]|e,n,\mu\rangle\langle e,n,\mu|\nonumber\\
&~&~+\hbar\left[(n+\mu)\omega-\frac{\omega_{0}}{2}-\frac{g^{2}n}{\Delta}\right]|g,n,\mu\rangle\langle
g,n,\mu|~~~~0<t\leq\tau_{1}\nonumber\\
H&=&\sum_{n,\mu}\hbar\left[(n+\mu)\omega+\frac{\omega_{0}}{2}\right]|e,n,\mu\rangle\langle
e,n,\mu|\nonumber\\
&~&~+\hbar\left[(n+\mu)\omega-\frac{\omega_{0}}{2}\right]|g,n,\mu\rangle\langle g,n,\mu|~~~~~~~~~~\tau_{1}<t\leq\tau_{1}+T\nonumber\\
H&=&\sum_{n,\mu}\hbar\left[(n+\mu)\omega+\frac{\omega_{0}}{2}+\frac{g_{2}^{2}(\mu+1)}{\Delta}\right]|e,n,\mu\rangle\langle
e,n,\mu|\nonumber\\
&~&~+\hbar\left[(n+\mu)\omega-\frac{\omega_{0}}{2}-\frac{g^{2}\mu}{\Delta}\right]|g,n,\mu\rangle\langle
g,n,\mu|
~~~~~~\tau_{1}+T<t\leq\tau_{1}+T+\tau_{2}.
\label{eq44}
\end{eqnarray}
A single atom initially in the superposition state $|\psi_{atom}\rangle=c_{e}|e\rangle+c_{g}|g\rangle$ is passed
through the cavities and the cavities have the fields in the coherent states $|\alpha\rangle$ and $\beta\rangle$, respectively.
 The state of the atom-cavity system at any time t is found to be,
\begin{equation}
|\psi_{ac}(t)\rangle=\sum_{n,\mu}A_{n,\mu}(t)|e,n,\mu\rangle+B_{n,\mu}(t)|g,n,\mu\rangle,
\end{equation}
where the coefficients acquire a time dependent phase as we are dealing with an interaction which is diagonal and
where 
\begin{equation}
\frac{A_{n,\mu}(0)}{c_{e}}=\frac{B_{n,\mu}(0)}{c_{g}}=
\frac{\alpha^{n}\beta^{\mu}}{\sqrt{n!\mu!}}\exp\left[-\left(\frac{|\alpha|^{2}+|\beta|^{2}}{2}\right)\right].
\label{coeff}
\end{equation}
 The state of the system after passing the atom through the cavities at time
$\tau(\tau>\tau_{1}+T+\tau_{2})$ is given by,
\begin{eqnarray}
|\psi_{ac}(\tau)\rangle=\sum_{n,\mu}A_{n,\mu}(0)&~&\exp\left[-i\left\{(n+\mu)\omega+\frac{\omega_{0}}{2}\right\}
\tau
-\frac{ig_{1}^{2}(n+1)\tau_{1}}{\Delta}-\frac{i
g_{2}^{2}(\mu+1)\tau_{2}}{\Delta}\right]|e,n,\mu\rangle\nonumber\\
+\sum_{n,\mu}B_{n,\mu}(0)&~&\exp\left[-i\left\{(n+\mu)\omega-\frac{\omega_{0}}{2}\right\}
\tau
+\frac{ig_{1}^{2}n\tau_{1}}{\Delta}+\frac{i
g_{2}^{2}\mu\tau_{2}}{\Delta}\right]|g,n,\mu\rangle.
\label{eq50}
\end{eqnarray}
Which on using the values of $A_{n,\mu}(0)$ and $B_{n,\mu}(0)$ leads to the compact result
\begin{eqnarray}
|\psi_{ac}(\tau)\rangle&=&c_{e}\exp\left\{-\frac{i}{\Delta}(g_{1}^{2}\tau_{1}+g_{2}^{2}\tau_{2})\right\}
|e,\alpha_{0}
e^{-\frac{ig_{1}^{2}\tau_{1}}{\Delta}},\beta_{0} e^{-\frac{ig_{2}^{2}\tau_{2}}{\Delta}}\rangle\nonumber\\
&+&c_{g}\exp(i\omega_{0}\tau) |g,\alpha_{0}
e^{\frac{ig_{1}^{2}\tau_{1}}{\Delta}},\beta_{0}e^{\frac{ig_{2}^{2}\tau_{2}}{\Delta}}\rangle,
\label{eq51}\\
\alpha_{0}&=&\alpha e^{-i\omega\tau},\beta_{0}=\beta e^{-i\omega\tau} .\nonumber 
\end{eqnarray}
If we choose the interaction times such that
$g_{1}^{2}\tau_{1}/\Delta=g_{2}^{2}\tau_{2}/\Delta=\pi/2$, and $c_{g}=c_{e}=1/\sqrt{2}$
and setting $i\alpha_{0}=\alpha^{'}, i\beta_{0}=\beta^{'}$, the state $(\ref{eq51})$ becomes
\begin{equation}
|\psi_{ac}(\tau)\rangle=\frac{1}{\sqrt{2}}\left\{|g,\alpha^{'},\beta^{'}\rangle e^{i\omega_{0}\tau}
-|e,-\alpha^{'},-\beta^{'}\rangle\right\}.
\label{eq52}
\end{equation}
If we detect the atom after passing through the cavities in the state
$c^{'}_{g}|g\rangle+c^{'}_{e}|e\rangle$, then the
state of the field is  reduced to,
\begin{equation}
|\psi_{cavities}\rangle=\frac{c^{'*}_{g}}{\sqrt{2}}
|\alpha^{'},\beta^{'}\rangle e^{i\omega_{0}\tau}
-\frac{c^{'*}_{e}}{\sqrt{2}}|-\alpha^{'},-\beta^{'}\rangle .
\label{eq53}
\end{equation}
The state $(\ref{eq53})$ is the entangled state of two coherent fields.
\section{conclusions}
Before we conclude, we mention that cavities are not absolutely essential for
doing Ramsey interferometry with quantized fields, though the usage of cavities
results in the enhancement of atom-photon interaction. We could, for example, imagine
the usage of the correlated photons produced by a down converter for doing
Ramsey interferometry though such an interaction would be quite weak. A possible
arrangement is shown in the Fig.\ref{fig3}.
 The final results for the interference pattern can be obtained  in a way similar to the
derivation of the result $(\ref{eq5})$. The visibility of the interference
pattern depends on the characteristics of the beam splitter and correlations
between the down converted photons. Note that the arrangement of the
Fig.\ref{fig3} generally makes 
$\langle a_{1}^{\dag}a_{2}\rangle \neq0$~. For the input state $|1,1\rangle $ the output 
state is \cite{15},
\begin{eqnarray}
|\psi_{out}\rangle =
(|t|^{2}-|r|^{2})|1,1\rangle+i\sqrt{2}|r||t|(|2,0\rangle+|0,2\rangle),~~~ |t|^{2}=1-|r|^{2},
\end{eqnarray}
where $|r|^{2}$ is the reflectivity of the beam splitter. For this state output
modes are correlated i.e.
\begin{equation} 
\langle a_{1}^{\dag}a_{2}\rangle=-2i(|t|^{2}-|r|^{2})|r||t|\neq0~,
\end{equation}
as long as $|r|\neq |t|$~. We note in passing that the problem of weak interaction between the atom and single
photon could possibly be overcome by
considering a collective system such as a Bose condensate \cite{bec}. The two Ramsey zones can be realised in the free
expansion of a condensate. The interaction in such a case would be enhanced by the 
square root of the density of atoms.

In conclusion we have discussed in detail the theory of Ramsey interferometry with quantized
fields. The interference is very sensitive to the quantum statistics of the
fields in the two Ramsey zones. We have derived general conditions for
interference to occur. We have shown how an analog of Hanbury-Brown Twiss
photon-photon correlation interferometry can be used to discern a variety of
interference effects even in situations where the single atom detection
probabilities do not exhibit interferences. We have demonstrated atoms acting as a mediator for
photon-photon interaction between two cavities and entanglement can be transfered
from fields to atoms. We have generated entangled state of two and
three photons by passing two and three atoms through the cavities. Further we have shown the generation of the
 entangled state of two coherent fields $|\alpha,\beta\rangle+|-\alpha,-\beta\rangle$ by using cavities in
 dispersive limit.

\newpage
\begin{figure}
\caption{A schematic arrangement for Ramsey interferometry with 
quantized fields. Each classical Ramsey zone is replaced by a cavity. There is 
a phase change between two cavities as $|e\rangle\rightarrow 
e^{-i\phi_{e}}|e\rangle$ and $|g\rangle\rightarrow e^{-i\phi_{g}}|g\rangle$.}
\label{fig1}
\end{figure}

\begin{figure}
\caption{Interference fringes in the probability of detecting a single 
atom in the excited state when the atom is initially in the
ground state for quantized (solidlines) and classical (dashedlines) fields. The
parameters are (a)
$g\tau=\pi, \Delta/g=10, \phi=0, V_q=0.68$  (b) $\Delta=0, g\tau=\pi/8 ,V_q=0.96$  (c)
$\Delta=0, g\tau=\pi/4, V_q=0.14E-01$ and  (d) $\Delta=0, g\tau=\pi/2, V_q=0.16$. The common parameters
for above graphs are $|\alpha|^{2}=5$, $\tau_{1}=\tau_{2}=\tau$, $g_{1}=g_{2}=g, V_c=1.00$ . $V_c , V_q$ are
the visibilities for classical fields and quantized fields.}
\label{fig5}
\end{figure}

\begin{figure}
\caption{The visibility of interference fringes vs.
$|\alpha|$ for weak classical fields (solidline) and for a nonclassical state (dashedline)
$\frac{1}{\sqrt{1+|\alpha|^{2}}}(|0\rangle+\alpha|1\rangle)$, with parameters
$g\tau_1=g\sqrt{2}\tau_2=\pi/2$ and $\Delta=0$.}  
\label{graph1}
\end{figure}

\begin{figure}
\caption{Case when both Ramsey zones are created inside a single cavity}
\label{fig2}
\end{figure}

\begin{figure}
\caption{An alternate scheme for Ramsey interferometry with quantized fields. 
The input fields $a_{0}$ and $b_{0}$ could be the outputs of a down converter.}
\label{fig3}
\end{figure}

\newpage
\begin{references}

\bibitem{1} N. F. Ramsey, Phys. Rev. {\bf78},
695, (1950).
\bibitem{2}
M. M. Salour and C. Cohen-Tannoudji, Phys. Rev. Lett. {\bf38}, 757, (1977);
     J. C. Bergquist, S. A. Lee, and J. L. Hall, 
     ibid. {\bf38}, 159 (1977);
M. M. Salour,
     Rev. Mod. Phys. {\bf50}, 667 (1978).
\bibitem{3}J. M. Raimond, M. Brune and S. Haroche, Rev. Mod. Phys. {\bf 73},
565 (2001). 
\bibitem{4}G. Nogues, A. Rauschenbeutel, S. Osnaghi, M. Brune, 
J. M. Raimond and S. Haroche, Nature(London) {\bf400}, 239, (1999).
\bibitem{5}P. Bertet, S. Osnaghi, A. Rauschenbeutel, G. Nogues, A. Auffeves, M.
Brune, J. M. Raimond and S. Haroche, Nature(London) {\bf411}, 166, (2001), for the theoretical proposal
see S. B. Zheng, Opt. Commun. {\bf173}, 265, (2000). Note that quantum eraser concept was proposed by M. O. Scully and
K. Dr$\ddot{u}$hl, Phys. Rev. A {\bf25}, 2208, (1982). 
\bibitem{berman}For extensive reviews on cavity quantum electrodynamics, see G. Raithel, C. Wagner, H. Walther, L.M. Narducci and M.O. Scully,
in {\bf Cavity Quantum Electrodynamics}, edited by P. R. Berman (Academic, Boston,
1994), p. 57; S. Haroche and J. M. Raimond, ibid. p. 123; H. J. Kimble, ibid. p. 203 ;
P. Meystre in {\bf Progress in Optics},edited by E. Wolf (Elsevier, Amsterdam, 1992) Vol.{\bf30}
 261; B. G. Englert etal Fortschr. Phys.{\bf46}, 900 (1998).
\bibitem{6} M. O. Scully, H. Walther, G. S. Agarwal, Tran Quang and W. Schleich,
Phys. Rev. A  {\bf44}, 5992 (1991);
R. J. Brecha, A. Peters, C. Wagner
 and H. Walther, ibid. {\bf46}, 567 (1992). 
\bibitem{7}N. F. Ramsey, Phys. Rev. A {\bf48}, 80 (1993).
\bibitem{9}M. O. Scully, B. G. Englert and H. Walther, Nature(London)
{\bf351}, 111 (1991);
B. G. Englert, M. O. Scully and H. Walther, ibid. {\bf375}, 367 (1995).
\bibitem{hariharan}A number of very interesting optical interference experiments have been reported with fields at
single photon level, P. Hariharan and B. C. Sanders in {\bf Progress in Optics} 
(North-Holland, Amsterdam, 1996) Vol.{\bf36}, p.49.
\bibitem{prl1}A. I. Lvovsky and J. Mlynek, Phys. Rev. Lett. {\bf88}, 250401 (2002).
\bibitem{prl2}K. J. Resch, J. S. Lundeen and A. M. Steinberg, Phys. Rev. Lett. {\bf88},113601 (2002).
\bibitem{surendra}Y. J. Lu and Z. Y. Ou, Phys. Rev. Lett. {\bf88}, 023601 (2002); H. Deng, D. Erenso, R. Vyas and S.
Singh ibid {\bf86}, 2770 (2001).
\bibitem{10}For a preliminary discussion see, G. S. Agarwal and M. O. Scully, Phys.
Rev. A {\bf53}, 467, (1996). 
\bibitem{scully}M. O. Scully and M. S. Zubairy, {\bf Quantum Optics}, (Cambridge
University Press, Cambridge 1997), Sec 20.4.
\bibitem{8}Conditional measurements on successive atoms passing through a single high quality cavity have led to
the successful generation of trapped and Fock states of radiation field. M. Weidinger, B. T. H. Varcoe, R. Heerlein and H. Walther,
Phys. Rev. Lett. {\bf82}, 3795 (1999);
B. T. H. Varcoe, S. Brattke, M. Weidinger and H. Walther, Nature(London) {\bf403}, 
743 (2000);
S. Brattke, B. T. H. Varcoe and H. Walther, 
Phys. Rev. Lett. {\bf86}, 3534 (2001).
\bibitem{11}Entanglement resulting from the process of detection is being discussed extensively,
J. M. Raimond, M. Brune and S. Haroche, Rev. Mod. Phys. {\bf 73},
565 (2001); 
L. M. Duan, M. D. Lukin, J. I. Cirac and P. Zoller, Nature(London) {\bf414}, 413 (
2001); G. S. Agarwal, J. von Zanthier, C. Skornia and H. Walther,
Phys. Rev.A, {\bf65}, 053826 (2002).
\bibitem{12}Entanglement between two macroscopic samples using the passage
of photons has been observed by,
Brian Julsgaard, Alexander Kozhekin and Eugene S. Polzik, Nature(London) {\bf413}, 400
(2001).
\bibitem{mandel}
R. Ghosh and L. Mandel,
  Phys. Rev. Lett. {\bf59}, 1903 (1987) ;
Z. Y. Ou and L. Mandel,
    ibid. {\bf62}, 2941 (1989);     
Z. Y. Ou, X. Y. Zou, L. J. Wang and L. Mandel,
    ibid.  {\bf65}, 321 (1990).
\bibitem{15}L. Mandel and E. Wolf, in {\bf Optical Coherence and Quantum Optics}
(Cambridge University Press, 1995) p.646.
\bibitem{brune}J. M. Raimond, M. Brune and S. Haroche, Phys. Rev. Lett. {\bf79},
1964 (1997); M. Brune, E. Hagley, J. Dreyer, X. Maitre, A. Maali, C. Wunderlich, J.
M. Raimond and S. Haroche, ibid. {\bf77}, 4887, (1996).
\bibitem{bec}F. Minardi, C. Fort, P. Maddaloni, M. Modugno and M. Inguscio , Phys. Rev. Lett. {\bf87},
 170401, (2001), have recently reported ramsey fringes using Bose condensates by
 using semiclassical fields.
\end{references}
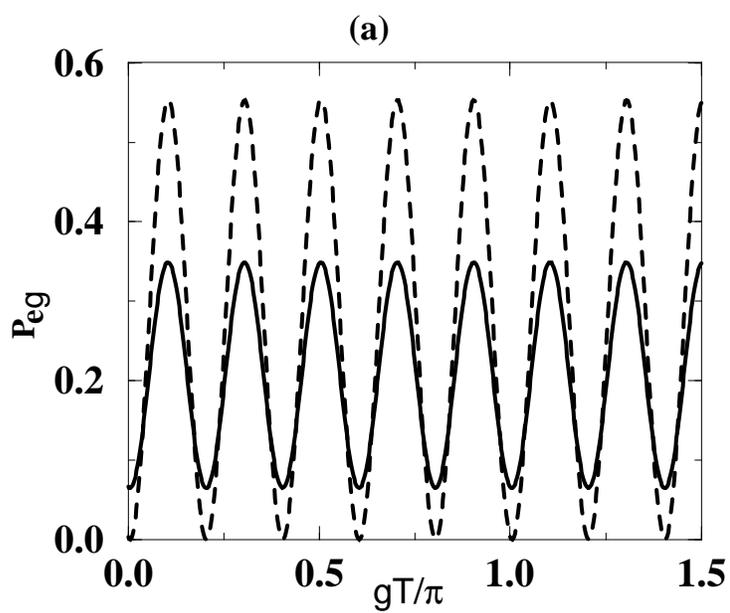
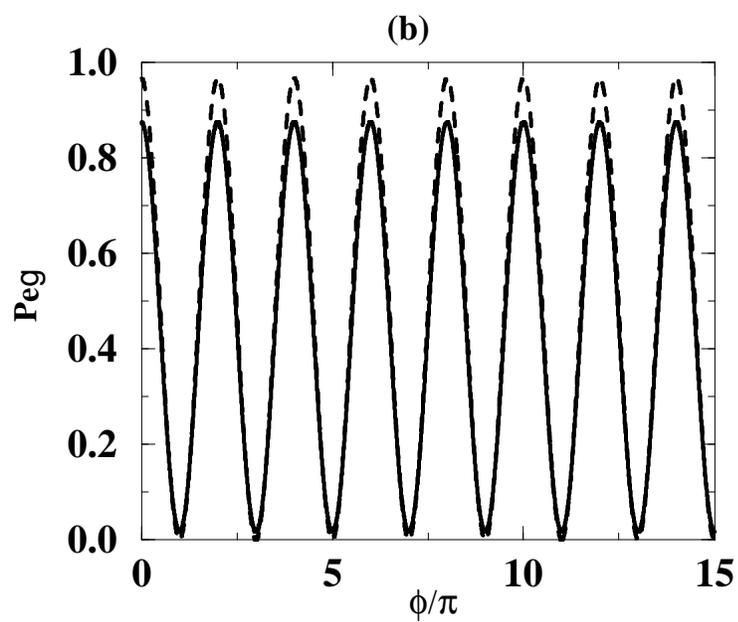
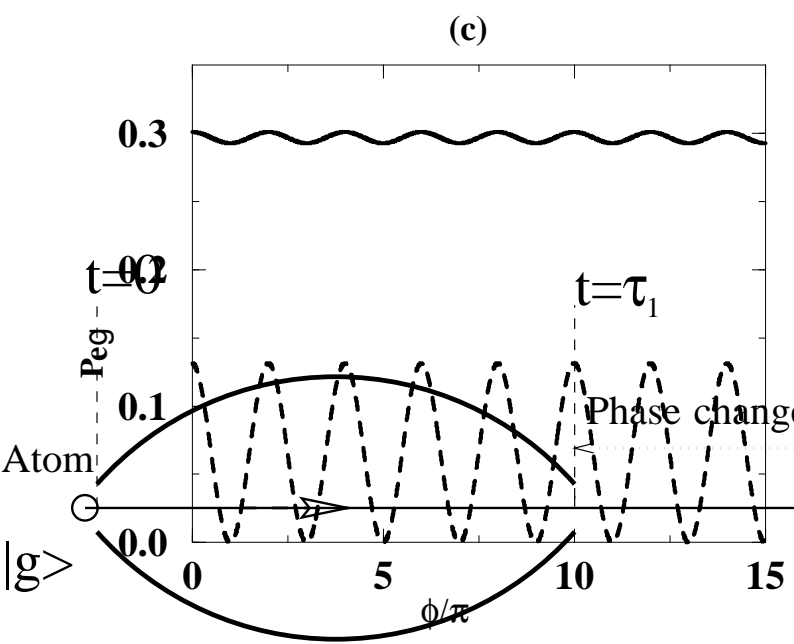
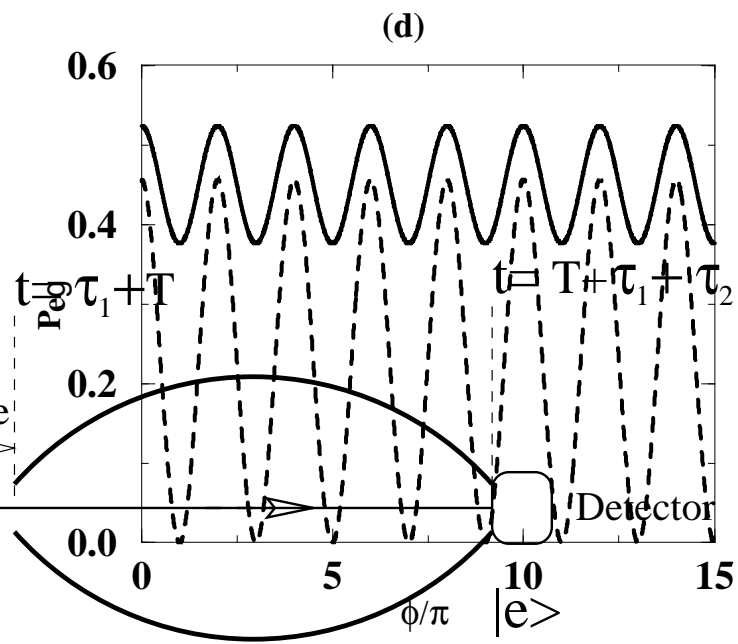
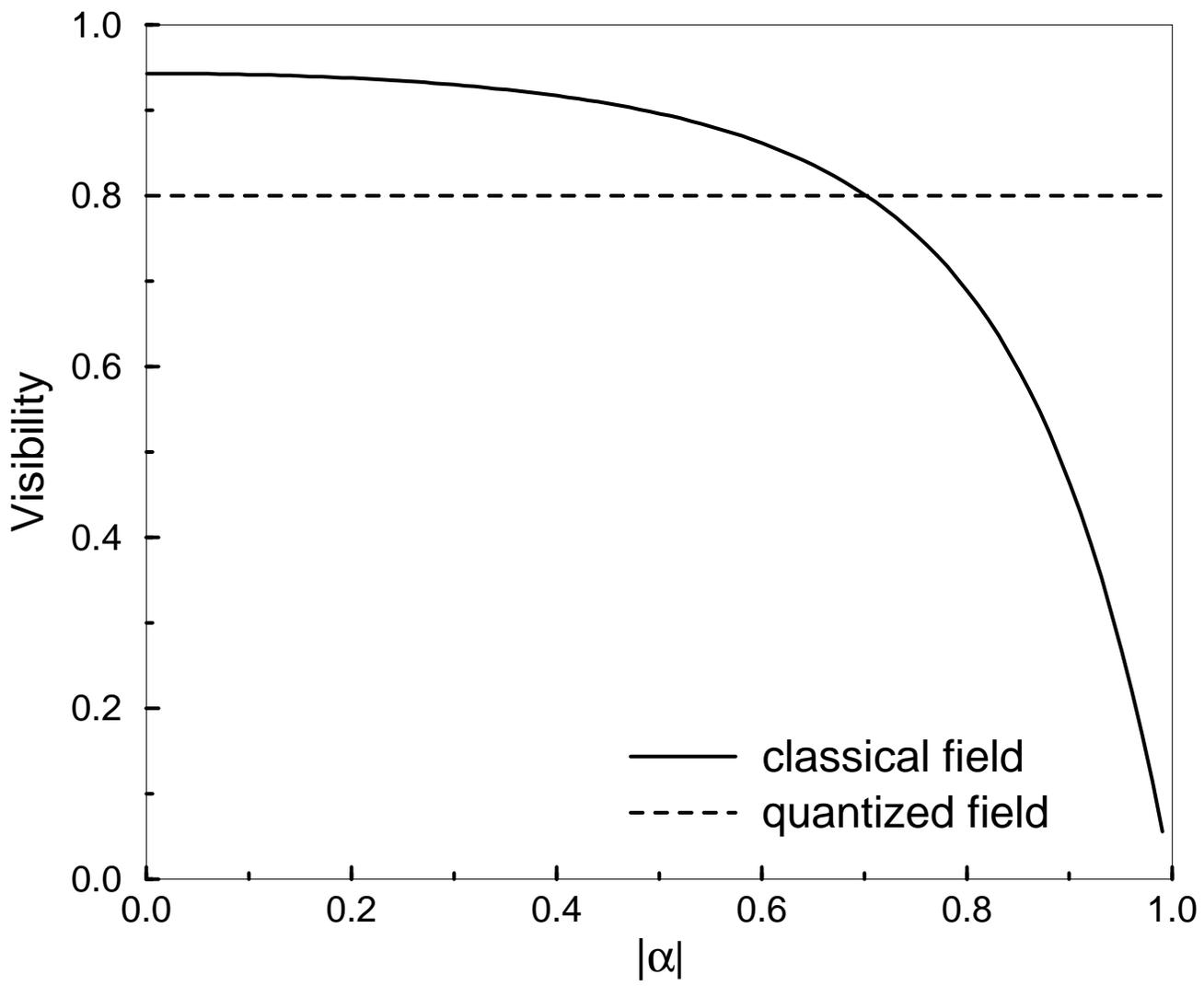
\end{document}